\documentclass[preprint, showpacs, reprintnumbers, amsmath, amssymb, superscriptaddress]{revtex4-2}

\usepackage{epsf}
\usepackage{graphicx}
\usepackage{sidecap}

 \usepackage{soul}
 \usepackage{array}
 \usepackage{amsmath}
 \usepackage{amssymb}
 \usepackage{dcolumn}
 \usepackage{epstopdf}
 \usepackage{bm}
 \usepackage{gensymb}
 \usepackage{amsmath}

\usepackage{color}
\usepackage{hyperref}
\usepackage{soul}
\sethlcolor{green}
\hypersetup{
    colorlinks=true,
    citecolor=red,
    linkcolor=red,
    filecolor=blue,   
    urlcolor=blue,
}
 
\def \beq {\begin{equation}}
\def \eeq {\end{equation}}
\begin{document}
 
\renewcommand{\figurename}{\textbf{Figure}}
\renewcommand{\thefigure}{{\textbf{\arabic{figure}}}}
 
\onecolumngrid

\begin{center}
 
  \textbf{\Large Observation of flat and weakly dispersing bands in a van der Waals semiconductor ${\mathrm{\textbf{Nb}}}_{3}{\mathrm{\textbf{Br}}}_{8}$ with breathing kagome lattice}\\[.2cm]
  Sabin~Regmi$^{1}$, Anup~Pradhan~Sakhya$^{1}$, Tharindu~Fernando$^{2}$, Yuzhou~Zhao$^{2}$, Dylan~Jeff$^{1,3}$, Milo~Sprague$^{1}$, Favian~Gonzalez$^{1,3}$, Iftakhar~Bin~Elius$^{1}$, Mazharul~Islam~Mondal$^{1}$, Nathan~Valadez$^{1}$, Damani~Jarrett$^{1}$, Alexis Agosto$^{1}$, Jihui~Yang$^{4}$, Jiun-Haw~Chu$^{2}$, Saiful~I.~Khondaker$^{1,3}$, Xiaodong~Xu$^{2}$, Ting~Cao$^{4}$, and Madhab~Neupane*$^{1}$\\[.2cm]
 {\itshape
    $^{1}$Department of Physics, University of Central Florida, Orlando, Florida  32816, USA\\
  	$^{2}$Department of Physics, University of Washington, Seattle, Washington 98195, USA\\
 	$^{3}$NanoScience and Technology Center, University of Central Florida, Orlando, Florida 32826, USA\\
 	$^{4}$Department of Materials Science and Engineering, University of Washington, Seattle, Washington 98195, USA}
 	\\ [.2cm]
$^*$Corresponding author: madhab.neupane@ucf.edu
\\[1cm]
\end{center}

\begin{abstract}
Niobium halides,  $\mathrm{Nb}_3X_8$ ($X = \mathrm{Cl}, \mathrm{Br}, \mathrm{I}$), which are predicted two-dimensional magnets, have recently gotten attention due to their breathing kagome geometry. Here, we have studied the electronic structure of $\mathrm{Nb}_3\mathrm{Br}_8$ by using angle-resolved photoemission spectroscopy (ARPES) and first-principles calculations. ARPES results depict the presence of multiple flat and weakly dispersing bands. These bands are well explained by the theoretical calculations, which show they have  $\mathrm{Nb}~d$ character indicating their origination from the $\mathrm{Nb}$ atoms forming the breathing kagome plane. This van der Waals material can be easily thinned down via mechanical exfoliation to the ultrathin limit and such ultrathin samples are stable as depicted from the time-dependent Raman spectroscopy measurements at room temperature. These results demonstrate that $\mathrm{Nb}_3\mathrm{Br}_8$ is an excellent material not only for studying breathing kagome induced flat band physics and its connection with magnetism, but also for heterostructure fabrication for application purposes.
\end{abstract}
\maketitle

Quantum materials with kagome lattice - a geometry of six triangles sharing the corners to form a hexagon within - in their crystal structure have been recently studied as the potential playgrounds for exploring the interplay among parameters such as geometry, topology, electronic correlations, magnetic and charge density orders \cite{Yu2012, Kiesel2012, Han2012, Wang2013, Kiesel2013, Mazin2014, Lin2018, Ye2018, Yin2019, YinTb2020, LinFeSn2020, Ghimire2020}. From the electronic structure point of view, a kagome lattice may support the presence of flat band, Dirac fermion, and saddle point with van Hove singularity. Angle-resolved photoemission spectroscopy (ARPES) \cite{Damascelli2003, Damascelli2004, Lv2019} has been successfully utilized to experimentally reveal some or all of these features in different kagome materials \cite{Lin2018, Ye2018, KangFeSn2020, LiuCoSn2020, KangCoSn2020, YinTb2020, Ortiz2020, Li2021, Dhakal2021, Peng2021, Kabir2022, Teng2022, Hu2022, Gu2022, Yang2022}. Majority of the reports have been on kagome systems with conventional kagome geometry, where bond lengths between the ions forming such geometry is equal so that the size of all the triangles is same. Kagome lattice can occur in a different geometry called the breathing kagome, where alternating triangles have different bond length between the constituent ions leading to different size \cite{Hanni2017, Ezawa2018, Bolens2019}. This difference may induce local electric dipole resulting in a ferroelectric order \cite{LiTi3X82021}.  Although it has been theoretically predicted that the breathing kagome systems can host intrinsically robust flat bands \cite{Bolens2019} and higher-order topology \cite{Ezawa2018}, the experimental studies of the breathing kagome systems for their electronic structure have been getting attention only  recently \cite{Tanaka2020, Sun2022, Regmi2022, Guo2022}.

Niobium halides $\mathrm{Nb}_3X_8$ ($X=\mathrm{Cl},\mathrm{Br},\mathrm{I}$) \cite{Magonov1993}, which possess breathing kagome plane formed by the $\mathrm{Nb}$ atoms, present themselves as material platforms to investigate the interplay of the breathing kagome geometry with magnetism and electronic correlations in both three- and two-dimensional limits. These compounds exhibit some intriguing attributes  that are advantageous for optoelectronic and nanodevice applications \cite{Jiang2017, Mortazavi2022}. They are moderate band gap semiconductors \cite{Sun2022, Oh2020} and in the monolayer form, they are predicted ferromagnet candidates \cite{Jiang2017, Conte2020, Cantele2022}. The introduction of additional layers to the monolayer is predicted to lead to an antiferromagnetic ordering \cite{Conte2020}. Importantly, because of very weak van der Waals coupling, they have very low exfoliation energies \cite{Mortazavi2022}. Therefore, obtaining the monolayers of these compounds is easily possible with mechanical exfoliation of the bulk crystals \cite{Sun2022, Regmi2022, Kim2017, Oh2020, Yoon2020}, which is beneficial for the fabrication of heterostructures. Recently, $\mathrm{Nb}_3\mathrm{Br}_8$ has been used to fabricate a heterostructure with $\mathrm{Nb}\mathrm{Se}_2$ to form a Josephson junction that can control the direction of current without the need of magnetic field \cite{Wu2022}. Bulk $\mathrm{Nb}_3\mathrm{Br}_8$ is in the singlet magnetic ground state at room temperature \cite{Pasco2019}. Although the density functional theory (DFT)-based computations have been performed to predict the band structure of the monolayer of this compound \cite{Jiang2017}, experimental demonstration of the electronic structure is still lacking.

In this paper, by means of ARPES measurements and supportive DFT computations, we report the electronic structure of $\mathrm{Nb}_3\mathrm{Br}_8$. The results of the ARPES measurements are consistent with the semiconducting nature of the material and the computed band structures well reproduce the experimental observations. Multiple flat and weakly dispersing bands are observed in the electronic band structure. The orbital-resolved calculations suggest these bands to have $\mathrm{Nb}~d$ character indicative of the origination from the breathing kagome plane of $\mathrm{Nb}$ atoms.  Moreover, through mechanical exfoliation of bulk crystal, a thin 6L sample of this material has been obtained and the stability of the sample in its ultrathin limit has been demonstrated through time-dependent room-temperature Raman spectroscopy measurements. This  study highlights $\mathrm{Nb}_3\mathrm{Br}_8$ as an excellent material candidate from both physics and application points of view by revealing the occurrence of flat band physics originating from breathing kagome lattice and by demonstrating an easy exfoliation of ultrathin sample and its stability at room temperature.

\begin{figure} [t!]
\centering
\includegraphics[width=0.75\textwidth]{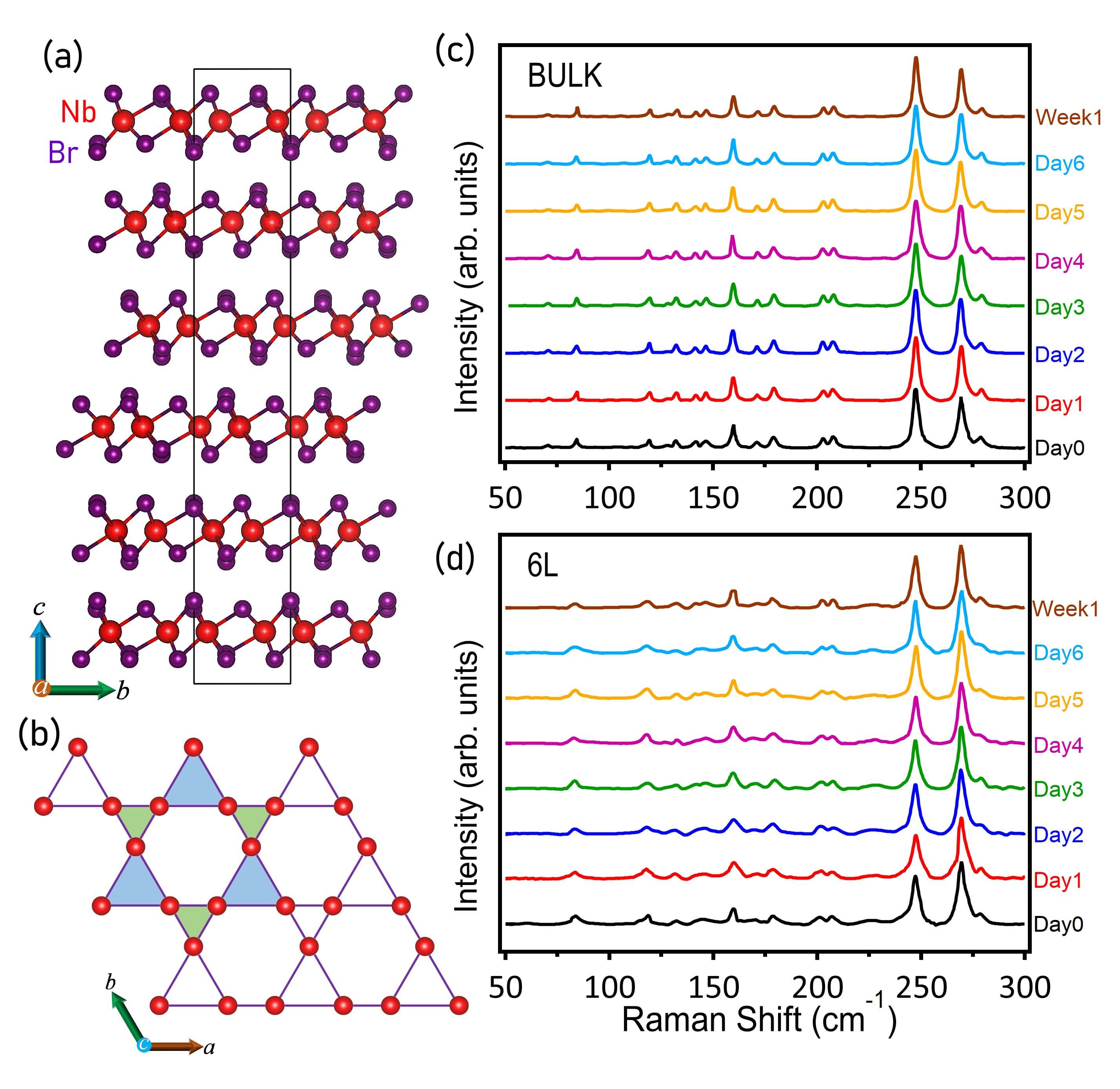}
\caption{Crystal structure and Raman spectroscopy measurements. (a) Crystal structure of ${\mathrm{Nb}}_{3}{\mathrm{Br}}_{8}$. Red and purple balls represent Nb and Br atoms, respectively. (b) Breathing kagome  plane of $\mathrm{Nb}$ atoms. Up triangles (light blue) and down triangles (light green) have different bond lengths. (c,d) Time dependent Raman spectra of bulk and 6L ${\mathrm{Nb}}_{3}{\mathrm{Br}}_{8}$, respectively.} 
\label{F1}
\end{figure}

High-quality single crystals of ${\mathrm{Nb}}_{3}{\mathrm{Br}}_{8}$ used for this study were grown by using the chemical vapor transport method. The crystal structure and the chemical composition were checked by using x-ray diffraction and energy dispersive x-ray spectroscopy. The ARPES studies on these crystals were carried out at the Stanford Synchrotron Radiation Lightsource endstation 5-2, which is equipped with a DA30 analyzer. DFT \cite{HohenbergKohn1964, KohnSham1965}-based first-principles computational results were implemented in the {\sc Vasp} package  with Projector Augmented Wave pseudopotential \cite{Kresse1996a, Kresse1996b, Kresse1999}. The mid band-gap is set to be the zero-energy level for comparison with the experimental results. Hubbard potential $U$ = 1~eV on $\mathrm{Nb}~d$  orbitals is used to address the effects of on-site Coulomb interactions. Details on the experimental and computational methods have been provided in the supplemental material (SM) section I \cite{SM}.

\begin{figure} [t!]
\includegraphics[width=0.85\textwidth]{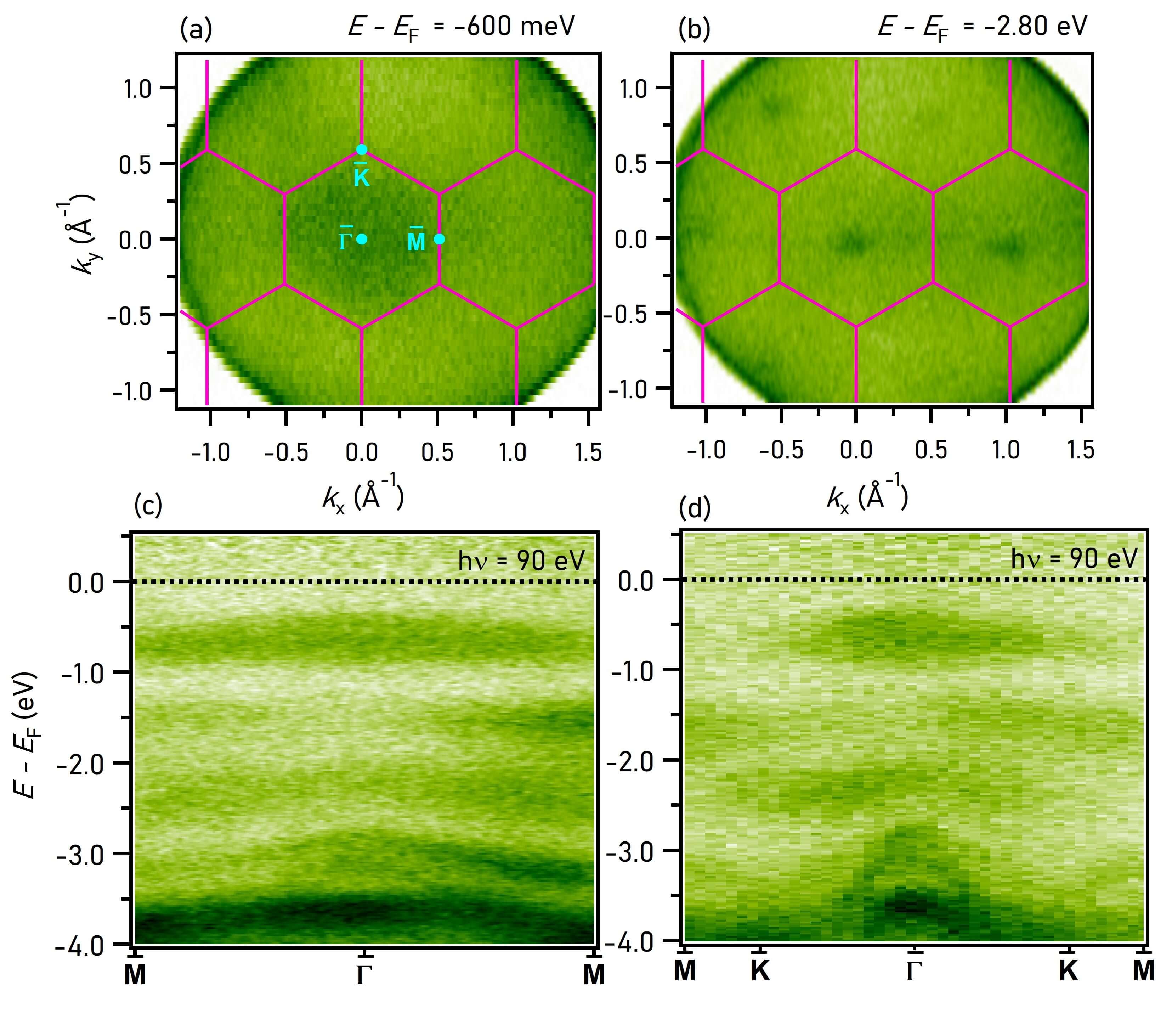}
\caption{Energy contours and band dispersion along different high-symmetry directions. (a,b) Energy contours measured by ARPES at the noted binding energies of $\mathrm{-600~meV}$ and $\mathrm{-2.8~eV}$, respectively. (c-d) Experimental band structures  along $\overline{\mathrm{M}}-\overline{\mathrm{\Gamma}}-\overline{\mathrm{M}}$ and $\overline{\mathrm{M}}-\mathrm{\overline{K}}-\overline{\mathrm{\Gamma}}-\mathrm{\overline{K}}-\overline{\mathrm{M}}$.}
\label{F2}
\end{figure}

${\mathrm{Nb}}_{3}{\mathrm{Br}}_{8}$ is a van der Waals layered material that crystallizes in rhombohedral space group $\mathrm{R\overline{3}m}$ (\# 166) with lattice parameters ${a}$ = ${b}$ = $\mathrm{7.080~\AA}$ and ${c}$ = $\mathrm{38.975~\AA}$ \cite{Magonov1993}. As shown in Fig.~\ref{F1}(a), a bulk unit cell is composed of six monolayers of ${\mathrm{Nb}}_{3}{\mathrm{Br}}_{8}$, where  the neighboring layers are connected through a very weak van der Waals interaction along the  crystallographic ${c}$-axis. In each layers, $\mathrm{Nb}$ atoms are arranged in a two-dimensional plane with breathing kagome geometry [Fig.~\ref{F1}(b)] and are sandwiched in between the bilayers of $\mathrm{Br}$ atoms on either sides. Because of the weak van der Waals coupling, the crystals are easily cleavable along the $\mathrm{(0001)}$ direction. The ${ab}$-plane orientation of the cleaved surface is indicated by the observation of sharp $\mathrm{(000l)}$ peaks in the single-crystal x-ray diffraction pattern [see Fig.~S1 in the SM \cite{SM}]. Importantly, bulk crystals can be easily exfoliated to ultrathin limits mechanically. We have thinned down the bulk crystal by using mechanical exfoliation to obtain a 6L ${\mathrm{Nb}}_{3}{\mathrm{Br}}_{8}$ [see Fig.~S2 in the SM \cite{SM}]. In Fig.~\ref{F1}(c), we present the results of the Raman spectroscopy measurements carried out on bulk and the 6L thin ${\mathrm{Nb}}_{3}{\mathrm{Br}}_{8}$. Tracking the evolution of the Raman spectra over time in Fig.~\ref{F1}(c,d), it is observed that the Raman modes do not experience shifts in peak frequency, changes in peak intensity, or peak broadening in either the bulk or 6L data. The absence of these changes in the Raman spectra signifies that the crystals are retaining good crystallinity throughout, speaking to the stability of the material in both bulk and thin layers. The fact that the crystal structure of ${\mathrm{Nb}}_{3}{\mathrm{Br}}_{8}$ is stable within a week’s time is promising for the future application of the material in technology or further research of its fundamental properties. Recent magnetic susceptibility study reports that  the bulk  ${\mathrm{Nb}}_{3}{\mathrm{Br}}_{8}$ undergoes a magnetic transition to a singlet state at around $\mathrm{382~K}$ \cite{Pasco2019}. Our magnetic susceptibility measurement [see Fig.~S1 in the SM \cite{SM}] shows an abrupt jump at a temperature of around $\mathrm{394~K}$, which is slightly off the reported transition temperature. 

 \begin{figure} [t!]
\includegraphics[width=0.85\textwidth]{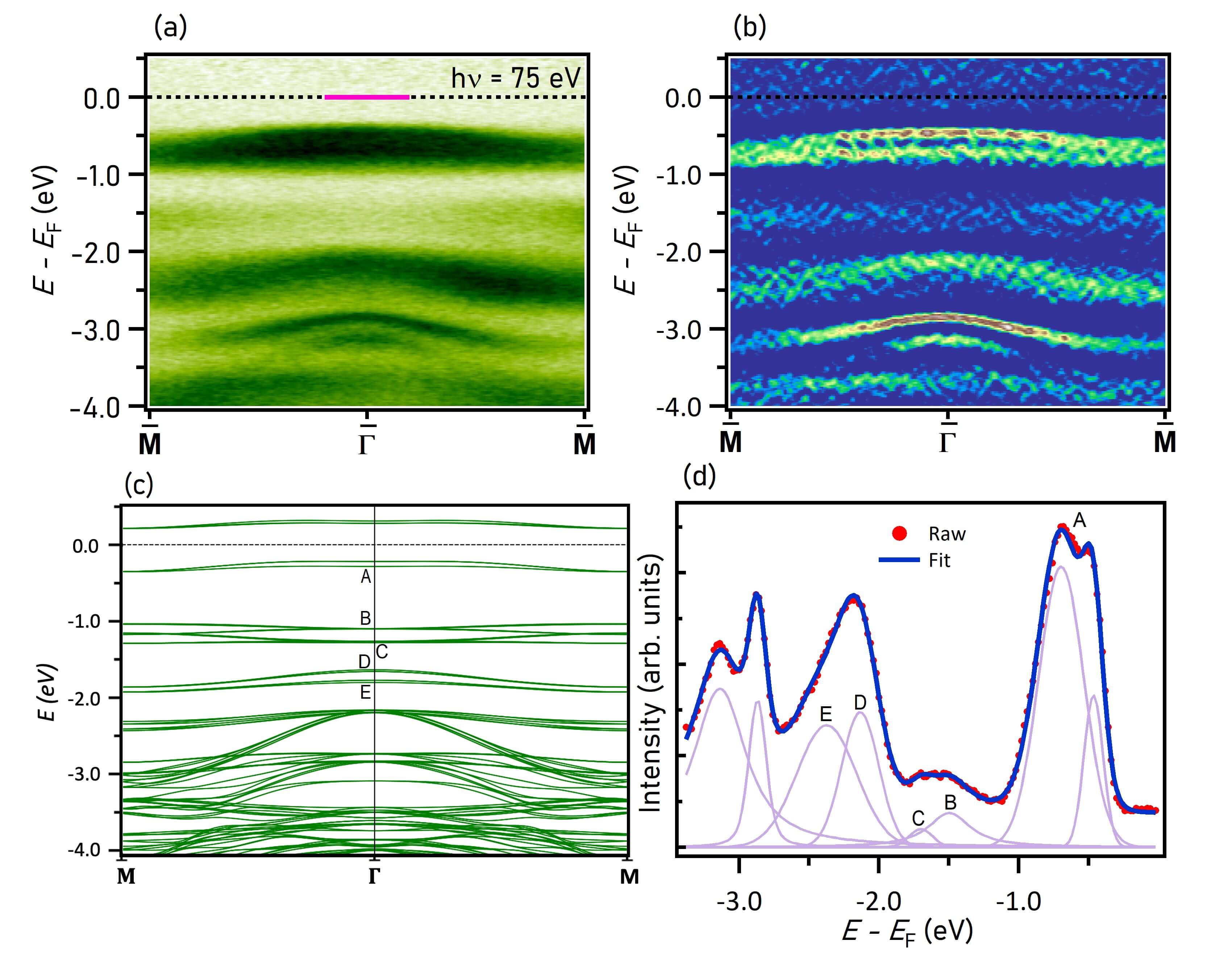}
\caption{Experimental and theoretical $\overline{\mathrm{M}}-\overline{\mathrm{\Gamma}}-\overline{\mathrm{M}}$ band dispersion. (a) Band dispersion along $\overline{\mathrm{M}}-\overline{\mathrm{\Gamma}}-\overline{\mathrm{M}}$ measured using incident photon energy of $\mathrm{75~eV}$  and (b) its second derivative plot. (c) Calculated band structure along $\mathrm{M}-\mathrm{\Gamma}-\mathrm{M}$.  (d) EDC integrated within a momentum window of ($\mathrm{-0.1~/\AA,~0.1~/\AA}$) represented by the magneta line in (a) and its Voigt fit.}
\label{F3}
\end{figure}

\begin{figure} [t!]
\centering
\includegraphics[width=0.60\textwidth]{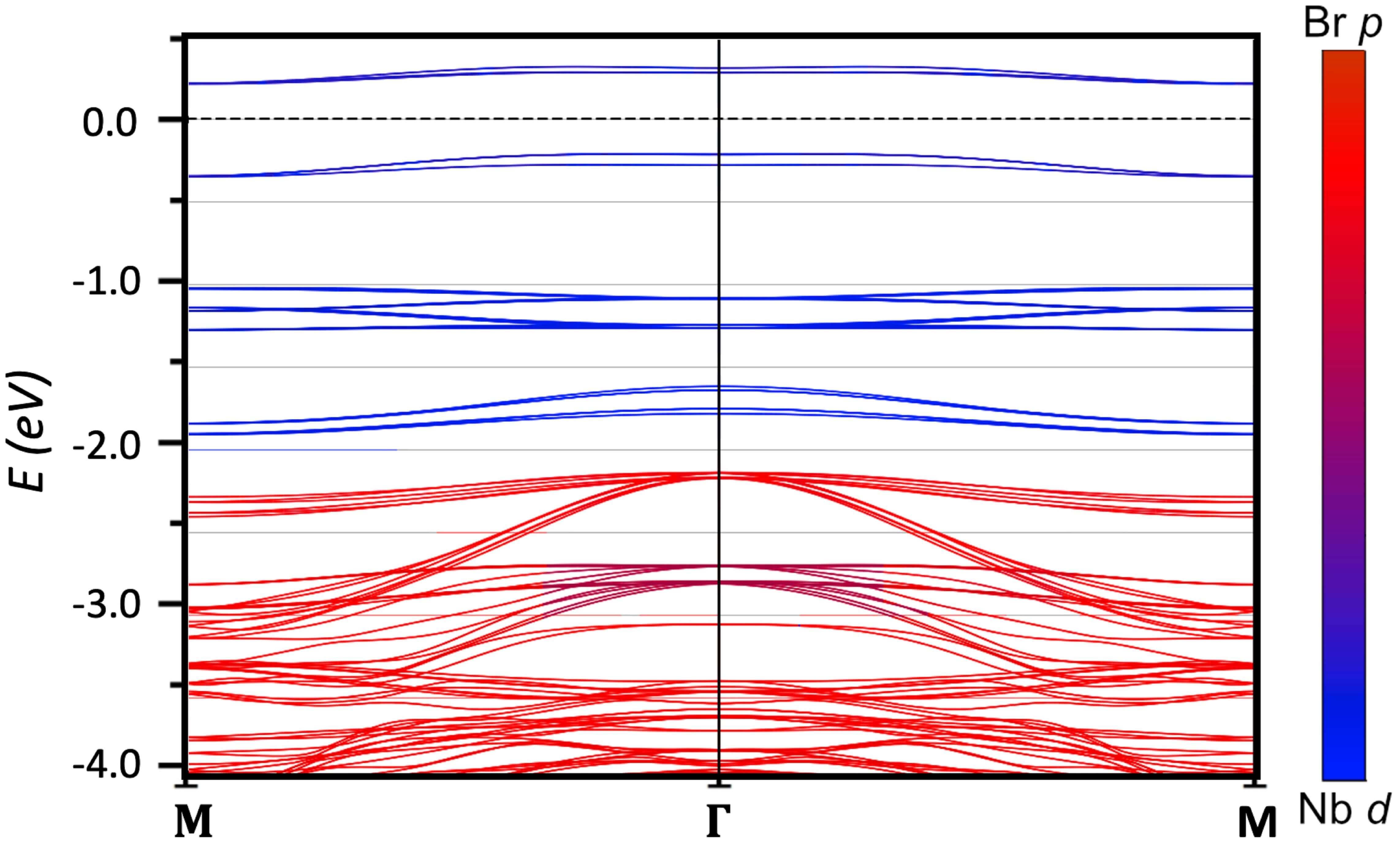}
\caption{Computed contribution of $\mathrm{Nb}~d$ and $\mathrm{Br}~p$ in the band structure along $\mathrm{M}-\mathrm{\Gamma}-\mathrm{M}$.}
\label{F4}
\end{figure}

The results of ARPES and accompanying first-principles band structure calculations are presented in Figs.~\ref{F2}-\ref{F4}, with an aim to unveil the electronic structure and the potential presence of  flat bands arising from the breathing kagome geometry in the crystal structure. The energy contours obtained from APRES measurement with $\mathrm{90~eV}$ incident photon energy  at binding energies of $\mathrm{-0.6~eV}$ and $\mathrm{-2.8~eV}$ are presented in Figs.~\ref{F2}(a) and \ref{F2}(b), respectively. Consistent with the semiconducting nature of the material, no photoemission signal is obtained at the Fermi level. While a clear photoemission intensity can be visualized in the $\mathrm{-0.6~eV}$ energy contour [Fig.~\ref{F2}(a)], a hexagonal pattern, typical of kagome systems, can be seen in the $\mathrm{-2.8~eV}$ energy contour [Fig.~\ref{F2}(b)]. In order to reveal the underlying band structure along different high-symmetry directions, we took dispersion maps along the $\overline{\mathrm{M}}-\overline{\mathrm{\Gamma}}-\overline{\mathrm{M}}$ and $\overline{\mathrm{M}}-\mathrm{\overline{K}}-\overline{\mathrm{\Gamma}}-\mathrm{\overline{K}}-\overline{\mathrm{M}}$ directions. The high-symmetry points are marked in Fig.~\ref{F2}(a). The band structures along these high-symmetry directions seem to reveal similar features. Similar results also have been obtained for different photon energies of the incident light [see Fig.~S3 in the SM \cite{SM}]. 

Figures~\ref{F3}(a-b) show energy-momentum dispersion along $\overline{\mathrm{M}}-\overline{\mathrm{\Gamma}}-\overline{\mathrm{M}}$ measured with a photon energy of $\mathrm{75~eV}$ and its second derivative plot, respectively. These plots show the presence of multiple flat and weakly dispersing bands within $\mathrm{-2.5~eV}$ binding energy. In Fig.~\ref{F3}(c), we present the calculated band structure along the $\mathrm{M}-\mathrm{\Gamma}-\mathrm{M}$ direction. The calculated band structure seems to reproduce the experimental observations quite well [See Fig.~S4 in the SM \cite{SM} for comparison of experimental and calculated band dispersion along $\overline{\mathrm{M}}-\mathrm{\overline{K}}-\overline{\mathrm{\Gamma}}-\mathrm{\overline{K}}-\overline{\mathrm{M}}$]. For the comparison with the experimental results, the mid-bandgap in the calculation has been set to zero. We can see that the energy positions of the bands in experimental and calculations results do not match. This discrepancy might arise due to various factors such as the underestimation of the bandgap by DFT or the sample not being perfectly stoichiometric. Nevertheless, the calculated band structure shows the presence of  bandsets - labeled A $\rightarrow$ E - within $\mathrm{2~eV}$ below the Fermi level.  Bandset A has two almost flat and parallel bands near the ${\mathrm{\Gamma}}$ point, among which the upper one seems to weakly disperse away from ${\mathrm{\Gamma}}$ before merging together near the $\mathrm{M}$ point. These bands can be seen in the dispersion map in Fig.~\ref{F3}(a) and better visualized in its second derivative [Fig.~\ref{F3}(b)].  Near the center of the BZ, these bands are almost flat and form a gap, which narrows down going towards the $\mathrm{M}$ point at which they seem to meet each other. Similarly, bandsets B and C both consist of two flat bands, one of which seems to separate and become dispersive going away from the $\mathrm{\Gamma}$ point. The bands within the bandsets B and  C are hard to distinguish in the experimental dispersion map as they form a wide spectrum of intensity that gives rise to a flat feature centered around $\mathrm{1.6~eV}$ [see Figs.~\ref{F3}(a)]. In Fig.~\ref{F3}(d), we present the energy distribution curve (EDC) integrated within a momentum window of ($\mathrm{-0.1~/\AA,~0.1~/\AA}$) represented by the magenta line in the raw dispersion map in Fig.~\ref{F3}(a). Several intensity peaks corresponding to the presence of bands can be observed. In particular, a two-peak feature centered around $\mathrm{-0.6~eV}$ can be observed that corresponds to the pair of  almost flat bands (bandset A). By fitting the raw EDC data with Voigt functions, we obtain that these almost flat bands in bandset A are located around $\mathrm{-0.46~eV}$ and $\mathrm{-0.70~eV}$ binding energies, respectively. Another broad but much suppressed intensity feature can be observed centered around $\mathrm{1.6~eV}$ binding energy in the EDC, which corresponds to the observed flat dispersion in Fig.~\ref{F3}(a). A good fitting is obtained by considering presence of two peaks located around $\mathrm{-1.5~eV}$ and $\mathrm{-1.7~eV}$ binding energies. For both bandsets B and C, as the flat band and the flat region of the other band dispersing away from $\mathrm{\overline{\Gamma}}$ merge together near the center of the BZ, they appear as these single peaks in the intensity plot within the experimental resolution. Fitting the EDC integrated within ($\mathrm{\overline{M}}$, $\mathrm{\overline{M}-0.1~\AA^{-1}}$), where these bands seem to get separated, shows the presence of a pair of bands in each of the bandsets B and C [see the Fig.~S5 in the SM \cite{SM}]. Below bandset C, there exist a relatively dispersive bandset D and a weakly dispersing bandset E, which are centered around $\mathrm{2.13~eV}$ and $\mathrm{2.37~eV}$ near the $\mathrm{\overline{\Gamma}}$ point in the experimental data [Fig.~\ref{F3}(d)]. 

The measurements at varying photon energies show that the aforementioned flat and weakly dispersing bands seem to have fairly similar dispersion irrespective of the choice of the photon energy indicative of their origination from two-dimensional plane [see the Fig.~S3 in the SM \cite{SM} for photon energy dependent measurements]. To understand if the origination of the flat and dispersing bands is from the breathing kagome plane of $\mathrm{Nb}$ atoms, we plot in Fig.~\ref{F4} the contribution of $\mathrm{Nb}~d$ and $\mathrm{Br}~p$ in the electronic structure along $\mathrm{M}-\mathrm{\Gamma}-\mathrm{M}$. While the dispersive bands deep below the Fermi level have $\mathrm{Br}~p$ character, it can be seen that the frontier flat and weakly dispersing bands discussed above have $\mathrm{Nb}~d$ character [see Fig.~S8 in the SM \cite{SM} for individual $\mathrm{Nb}~d$ orbitals' contribution]. This confirms that the $\mathrm{Nb}$ atoms, which configure themselves in the breathing kagome geometry, give rise to these flat and weakly dispersing bands. $\mathrm{Nb}_3\mathrm{Br}_8$ is a predicted ferromagnet in its monolayer form \cite{Jiang2017}. To see how the band structure in the two-dimensional form will compare with the band structure of the bulk, we computed ferromagnetic band structure calculations for the monolayer [see Fig.~S6 in the SM \cite{SM}]. From the comparison, we anticipate that the frontier electronic band structure for thin samples and bulk will look similar overall. 

To conclude, the electronic structure of the niobium halide semiconductor $\mathrm{Nb}_3\mathrm{Br}_8$ has been studied by using ARPES and first-principles calculations. In its crystal structure, $\mathrm{Nb}$ atoms form a breathing kagome lattice with different $\mathrm{Nb}$\textendash$\mathrm{Nb}$ bond lengths in alternate triangles. Experimental dispersion maps reveal the presence of multiple weakly dispersing and flat bands, with the highest set of such bands occurring around $\mathrm{460}$ and $\mathrm{700~meV}$ below the Fermi level. The comparison with the first-principles computations show that these flat and weakly dispersing bands have $\mathrm{Nb}-d$ character suggesting their origination from the breathing kagome plane formed by the $\mathrm{Nb}$ atoms. Room temperature Raman measurements on a 6L  sample mechanically exfoliated from the bulk crystal depict the stability of the ultrathin sample suggesting $\mathrm{Nb}_3\mathrm{Br}_8$ as an excellent ground to study breathing kagome geometry induced flat band physics and its application.\\

M.~N. acknowledges the support from the National Science Foundation (NSF) CAREER award DMR-1847962, the NSF Partnerships for Research and Education in Materials (PREM) Grant DMR-2121953, and the Air Force Office of Scientific Research MURI Grant No. FA9550-20-1-0322. T.C., X.X., J.-H.C., J.Y., T.F., and Y.Z. acknowledge the support from UW Molecular Engineering Materials Center and a NSF Materials Research Science and Engineering Center (NSF MRSEC DMR-2308979). The computational work at UW was facilitated through the use of advanced computational, storage, and networking infrastructure provided by the Hyak supercomputer system and funded by the UW Molecular Engineering Materials Center (NSF MRSEC DMR-2308979). S.I.K. also acknowledges the support from the NSF PREM Grant DMR-2121953. This study uses the resources of the Stanford Synchrotron Radiation Lightsource (SSRL), SLAC National Accelerator Laboratory, which is supported by the U.S. Department of Energy, Office of Science, Office of Basic Energy Sciences under Contract No. DE-AC02-76SF00515. We acknowledge the beamline assistance from Dr. Makoto Hashimoto and Dr. Donghui Lu at SSRL.

 \clearpage

\setcounter{equation}{0}
\renewcommand{\theequation}{S\arabic{equation}}
\setcounter{figure}{0}
\renewcommand{\thefigure}{S\arabic{figure}}
\setcounter{section}{0}
\renewcommand{\thesection}{S\Roman{section}}
\setcounter{table}{0}
\renewcommand{\thetable}{S\arabic{table}}
\begin{center}
 \textbf{\Large \underline{Supplemental Material}}\\[0.5cm]
\end{center}
\begin{center}
\textbf{1. METHODS}\\
\textbf{1.1 Experimental techniques}\\
\textit{Crystal Structure and Sample Characterization} \end{center}
High-quality single crystals of ${\mathrm{Nb}}_3{\mathrm{Br}}_8$ were synthesized by utilizing chemical vapor transport method \cite{Pasco2019S}. $\mathrm{0.1084~g}$ high-purity $\mathrm{Nb}$ powder ($\mathrm{99.99\%}$, 325mesh from Alfa Aesar) and $\mathrm{0.6564~g}$ $\mathrm{Nb}\mathrm{Br}_5$ ($\mathrm{99.9\%}$ from Strem Chemicals Inc.) were sealed inside a quartz tube ($\mathrm{12.75~mm~OD} \times \mathrm{10.5~mm~ID} \times \mathrm{12~cm}$) under vacuum ($\mathrm{<20~mTorr}$). The tube was loaded into a dual zone furnace. Both heating zones were first ramped for $\mathrm{1~hour}$ to reach $\mathrm{200~\degree C}$, and kept at that temperature for $\mathrm{4~hours}$ before ramping for $\mathrm{10~hours}$ to reach $\mathrm{830~\degree C}$ on the precursor side and $\mathrm{785~\degree C}$ on the growth side. After $\mathrm{6~days}$ of growth, the precursor side and the growth side were cooled down for $\mathrm{8~hours}$ to reach $\mathrm{20~\degree C}$ and $\mathrm{350~\degree C}$, respectively and kept for $\mathrm{3~hours}$. This removes the excess precursor from the growth side. Then the furnace was naturally cooled down to obtain flat and shiny crystals. The crystal structure was examined using X-ray diffraction and the chemical compositions of the obtained crystals were verified using scanning electron microscopy and energy-dispersive X-ray spectroscopy analysis. X-ray diffraction pattern obtained on a single crystal of ${\mathrm{Nb}}_{3}{\mathrm{Br}}_{8}$ presented in FIG.~\ref{S1}(a) shows sharp peaks associated with (000l) peaks, which indicate the cleavability of the samples along the $ab$-plane.

\begin{figure}[h!]
\includegraphics[width=1\textwidth]{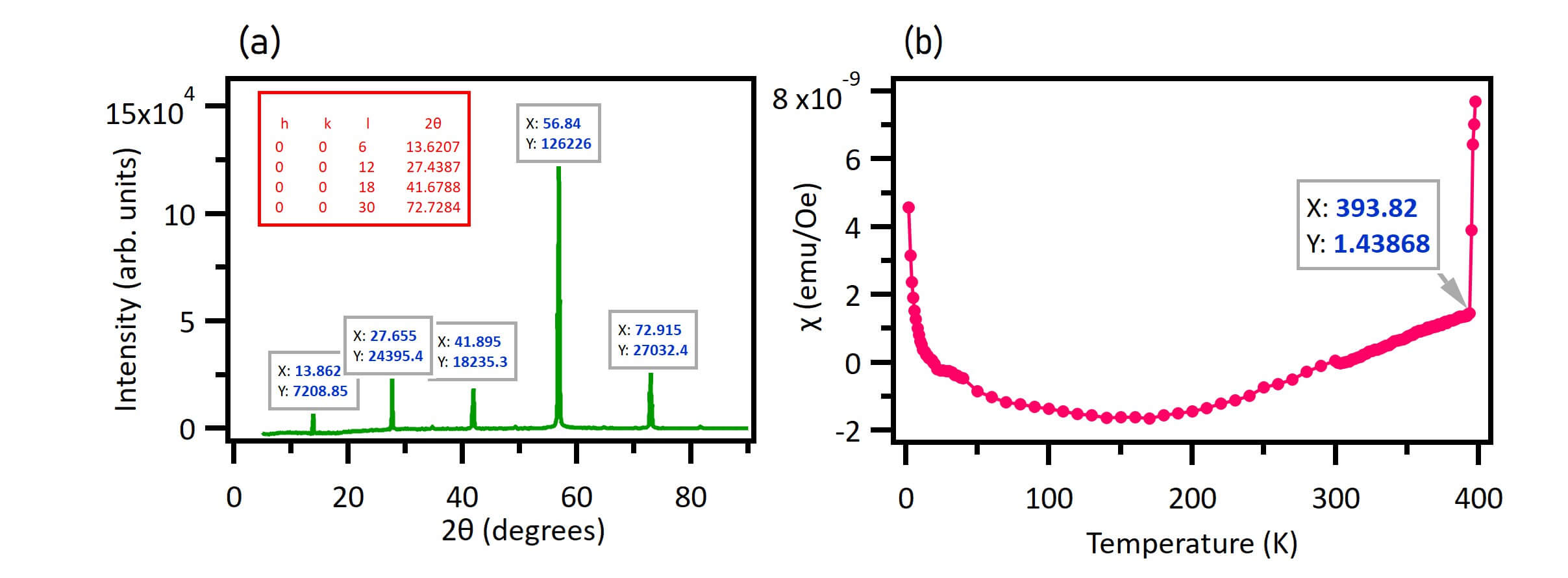}
\caption{Sample characterization. (a) Single crystal XRD. (b) Magnetic susceptibility measured as a function of temperature upto $\mathrm{400~K}$.}
\label{S1}
\end{figure}

Magnetic susceptibility ($\chi$) data were collected using a Quantum Design physical properties measurement system (PPMS) on single crystals using the vibrating sample magnetometer (VSM) option. The sample was mounted onto the in-plane quartz paddle sample holder using GE-7031varnish.  The same holder with only varnish was used as blank to subtract backgrounds. At each temperature, the magnetic moment was measured as function of magnetic field by sweeping field from $\mathrm{-5}$ to $\mathrm{5~T}$, and the slope between $\mathrm{-5}$ to $\mathrm{5~T}$ was fitted to extract the magnetic susceptibility of the sample at the temperature.  The VSM measurement was carried from $\mathrm{2}$ to $\mathrm{400~K}$. The obtained $\chi$ is presented in FIG.~\ref{S1}(b) as a function of temperature ($T$). $\chi(T)$ plot is similar to previously reported in the literature \cite{Pasco2019S}, with a jump at $\mathrm{\sim394~K}$ corresponding to the magnetic transition to a singlet ground state. The upturn at the low temperatures has been attributed to the defect spins \cite{Pasco2019S}.

\begin{figure}[h!]
\includegraphics[width=1\textwidth]{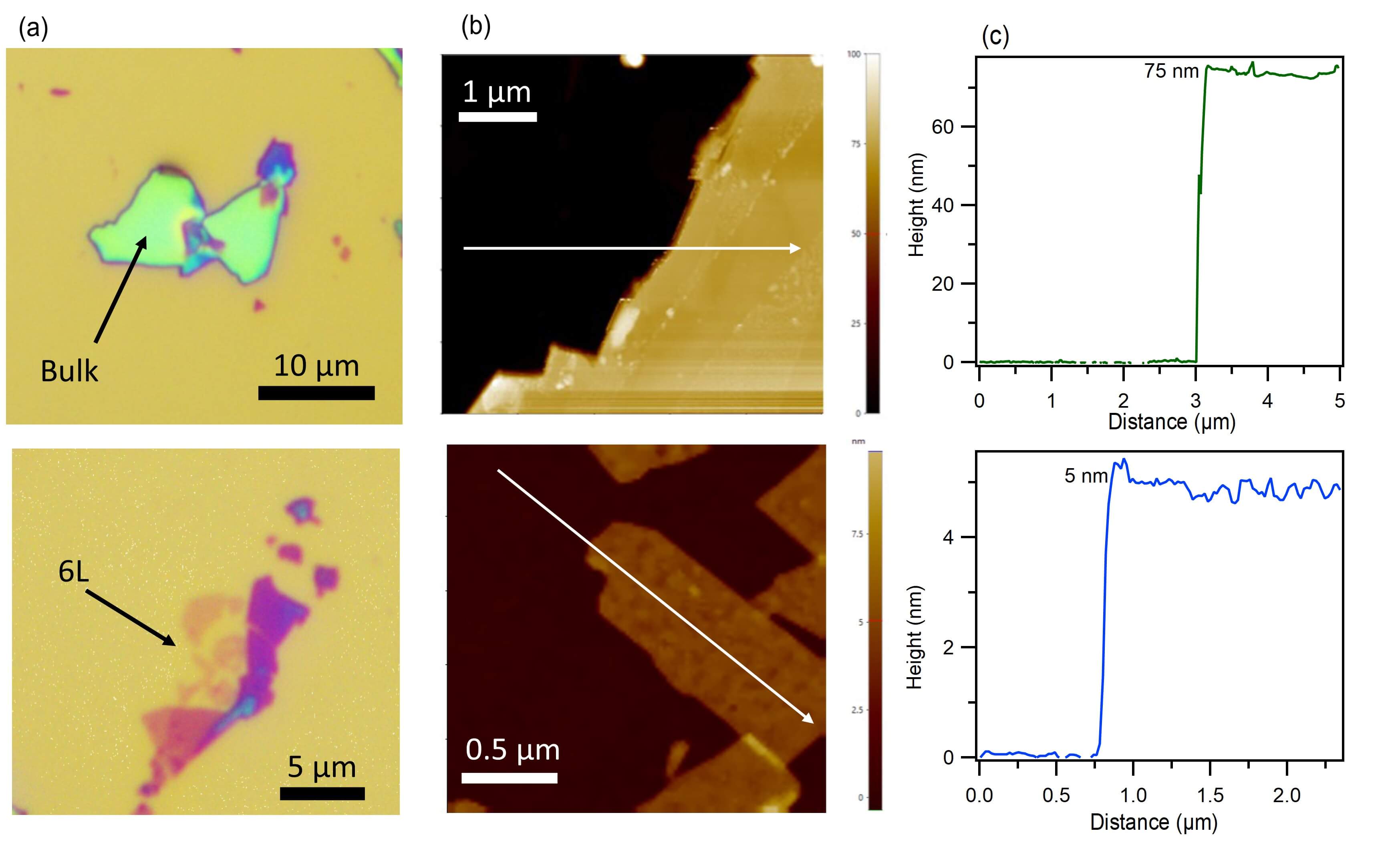}
\caption{Mechanical exfoliation. (a) Optical images of the flakes exfoliated from bulk ${\mathrm{Nb}}_{3}{\mathrm{Br}}_{8}$. (b) AFM images of the exfoliated flakes. (c) Height profile of the exfoliated flakes corresponding to the AFM images in (b).}
\label{S2}
\end{figure}

\begin{center}
\textit{Exfoliation and Raman spectroscopy measurements}
\end{center}
${\mathrm{Nb}}_{3}{\mathrm{Br}}_{8}$ flakes were mechanically cleaved from the bulk crystals and directly exfoliated onto silicon wafers with a $\mathrm{250~nm}$ $\mathrm{Si}\mathrm{O}_2$ capping layer using Nitto tape (SPV-224 PVC). Height measurements of the exfoliated ${\mathrm{Nb}}_{3}{\mathrm{Br}}_{8}$ flakes were performed via atomic force microscopy with Park Systems’ Park NX20 in non-contact mode. The optical and AFM images of the flakes have been presented in Figs.~\ref{S2}(a) and \ref{S2}(b), respectively. The corresponding height profile is presented in Fig.~\ref{S2}(c). Layer number was calculated by using the value reported in \cite{Wu2022S}.

Raman experiments were performed using a confocal microscope Raman spectrometer (LabRAM HR Evolution, Horiba Scientific) in a backscattering geometry. The laser was focused on the exfoliated ${\mathrm{Nb}}_{3}{\mathrm{Br}}_{8}$ flakes through a $\mathrm{100}\times$ objective with a spot size of $\mathrm{\sim 1~\mu m}$. The Raman emission was collected and dispersed by a $\mathrm{1800~gr/mm}$ grating, using a $\mathrm{532~nm}$ wavelength laser as the excitation source. The dispersed Raman signal is then detected by a CCD detector (Synapse-EM). All Raman measurements were taken in ambient conditions.

\begin{center}
\textit{ARPES measurements}
\end{center}
Synchrotron-based ARPES measurements were performed at the Stanford Synchrotron Radiation Lightsource (SSRL) endstation 5-2 equipped with a DA30 analyzer. Flat samples of ${\mathrm{Nb}}_{3}{\mathrm{Br}}_{8}$ were mounted on top of copper posts using silver epoxy paste. Ceramic posts were mounted on top of the flat samples again using the silver epoxy paste. So prepared samples were loaded into the ARPES chamber and then cleaved \textit{in situ} under ultra-high vacuum conditions maintained in the order of $\mathrm{10^{-11}~torr}$ to obtain fresh sample surface for ARPES measurements. Due to the semiconducting nature of the samples, the charging effects were excluded by taking measurements at a temperature of $\mathrm{300~K}$, which leads to the broadened data in the results of the ARPES measurements.

\begin{center}
\textbf{1.2 Computational details}
\end{center}
Density-functional theory (DFT) \cite{HohenbergKohn1964S, KohnSham1965S}-based computations were performed for bulk stable state (antiferromagnetic with alternate spin on alternate layers) and monolayer stable state (ferromagnetic). The computations were implemented in the {\sc Vasp} package with Projector Augmented Wave (PAW) pseudopotential \cite{Kresse1996aS, Kresse1996bS, Kresse1999S}. Perdew-Burke-Ernzerhof-type exchange-correlational functional \cite{PBE1996S} was used with van der Waals correction included in the D2 formalism \cite{Grimme2006S}. To account for the on-site coulomb interaction, Hubbard potential $U =\mathrm{1~eV}$ on $\mathrm{Nb}~d$ orbitals was used. For the bulk calculations, a unit cell with $a=b=\mathrm{7.080~\AA}$ and $c=38.975~\AA$ \cite{Magonov1993S} was used. For monolayer calculations, a single layer of $\mathrm{Nb}_3\mathrm{Br}_8$ was used with $c=\mathrm{25.000~\AA}$. $\mathrm{300~eV}$ energy cut-off was used for the plane-wave basis set and a $\Gamma$-centered $\mathrm{10 \times 10 \times 3}$ ($\mathrm{10 \times 10 \times 1}$) {\it k} mesh was employed for the bulk (monolayer) calculations. The mid bandgap was set as the zero energy level for comparison with the experimental results. For the plots, PyProcar \cite{Herath2020S} was used.

\begin{center}
\textbf{2. DISPERSION MAPS AT DIFFERENT PHOTON ENERGIES}
\end{center}

\begin{figure}[h!]
\includegraphics[width=1\textwidth]{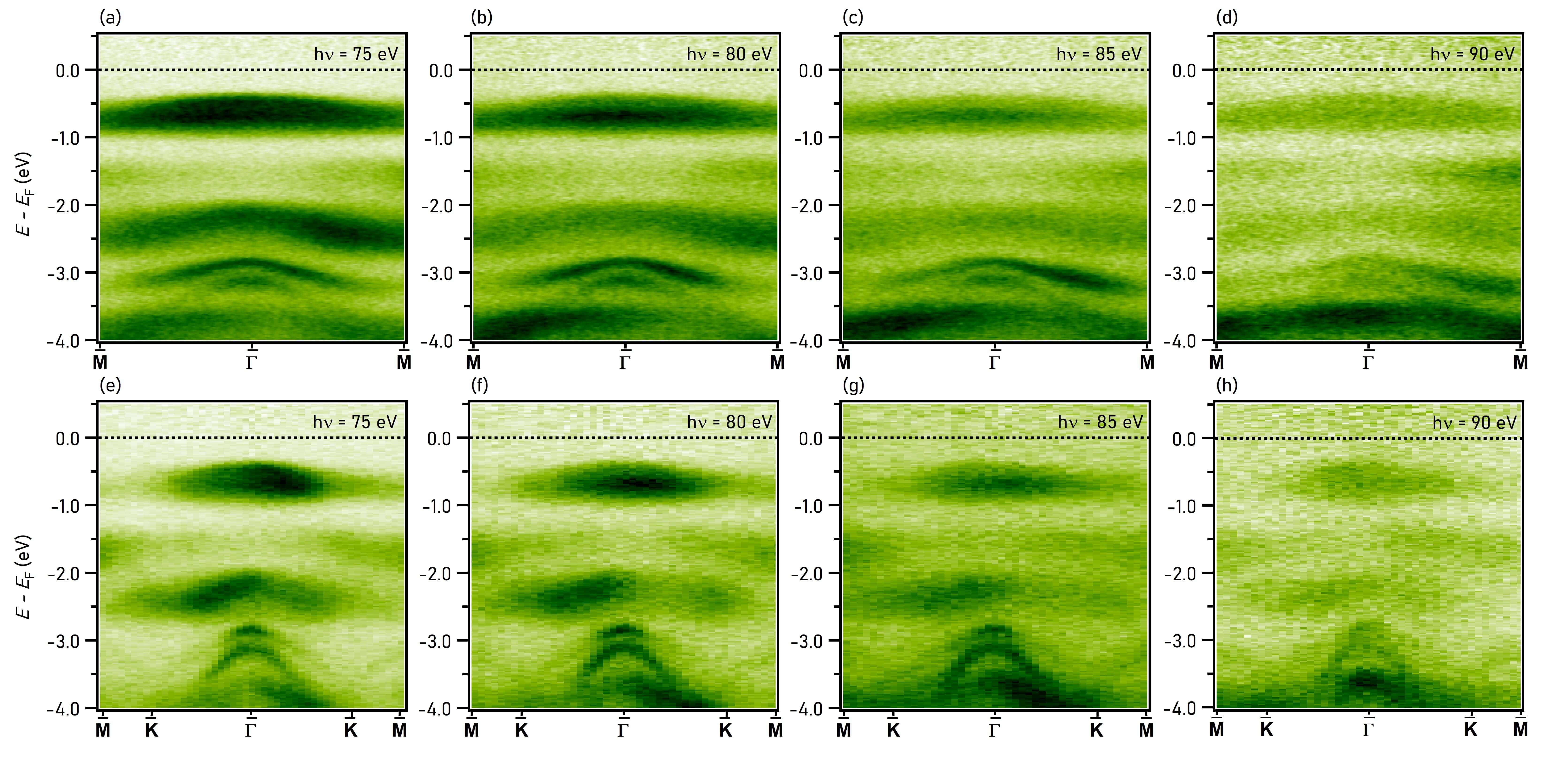}
\caption{Photon energy dependent measurements. (a-d) $\mathrm{\overline{M}}-\mathrm{\overline{\Gamma}}-\mathrm{\overline{M}}$ dispersion maps measured with incident photon energies as noted on each plots. (e-h) $\mathrm{\overline{M}}-\mathrm{\overline{K}}-\mathrm{\overline{\Gamma}}-\mathrm{\overline{K}}-\mathrm{\overline{M}}$ dispersion maps taken by using photon energies as noted on each plots.}
\label{S3}
\end{figure}
 In FIG.~\ref{S3}, experimentally obtained band dispersion along the $\mathrm{\overline{M}}-\mathrm{\overline{\Gamma}}-\mathrm{\overline{M}}$ and $\mathrm{\overline{M}}-\mathrm{\overline{K}}-\mathrm{\overline{\Gamma}}-\mathrm{\overline{K}}-\mathrm{\overline{M}}$ directions are presented. The overall electronic structure is similar for both the directions, where multiple flat and weakly dispersing bands can be seen within $\mathrm{2~eV}$ binding energy. Despite the variation in intensities, these flat and weakly dispersing bands seem to have similar dispersion for all photon energies. This is indicative of the two-dimensional origination of these bands.

\begin{figure}[h!]
\includegraphics[width=0.7\textwidth]{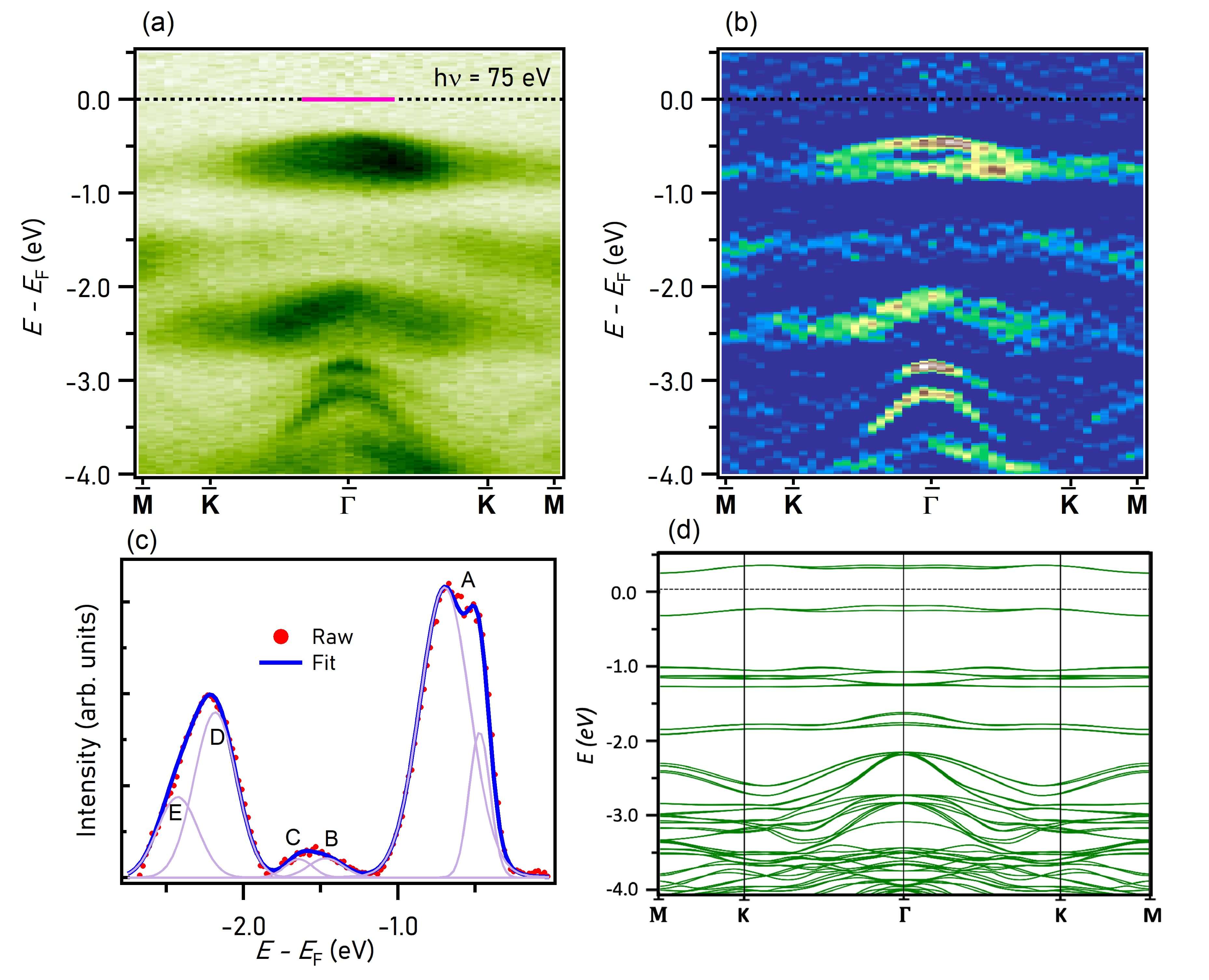}
\caption{$\mathrm{\overline{M}}-\mathrm{\overline{K}}-\mathrm{\overline{\Gamma}}-\mathrm{\overline{K}}-\mathrm{\overline{M}}$ band dispersion. (a) Experimental energy-momentum dispersion and its second derivative (b). (c) Integrated EDC within the momentum range represented by the magenta line in (a). (d) Calculated band structure along $\mathrm{M}-\mathrm{K}-\mathrm{\Gamma}-\mathrm{K}-\mathrm{M}$).}
\label{S4}
\end{figure}

\begin{center}
\textbf{3. EXPERIMENTAL AND CALCULATED BAND DISPERSION ALONG $\mathrm{\overline{M}}-\mathrm{\overline{K}}-\mathrm{\overline{\Gamma}}-\mathrm{\overline{K}}-\mathrm{\overline{M}}$}
\end{center}
In Fig.~\ref{S4}(a,b), we present the ARPES measured electronic band structure along the $\mathrm{\overline{M}}-\mathrm{\overline{K}}-\mathrm{\overline{\Gamma}}-\mathrm{\overline{K}}-\mathrm{\overline{M}}$ direction and its second derivative plot, respectively. An integrated energy distribution curve (EDC) for momentum window represented by the magenta line in \ref{S4}(a) is shown in Fig.~\ref{S4}(c). Peaks corresponding to bandsets A $\rightarrow$ E can be observed. Around the region where the bands within A are almost flat and well separated around the $\mathrm{\overline{\Gamma}}$ point, a two peak feature can be resolved. For B $\rightarrow$ E, a single wide peak is observed due to bands within these bandsets being almost merged together near  $\mathrm{\overline{\Gamma}}$. The DFT-calculated band structure along the $\mathrm{M}-\mathrm{K}-\mathrm{\Gamma}-\mathrm{K}-\mathrm{M}$, presented in Fig.~\ref{S4}(d), shows excellent agreement with these experimentally  observed results.

\begin{figure}[h!]
\includegraphics[width=1\textwidth]{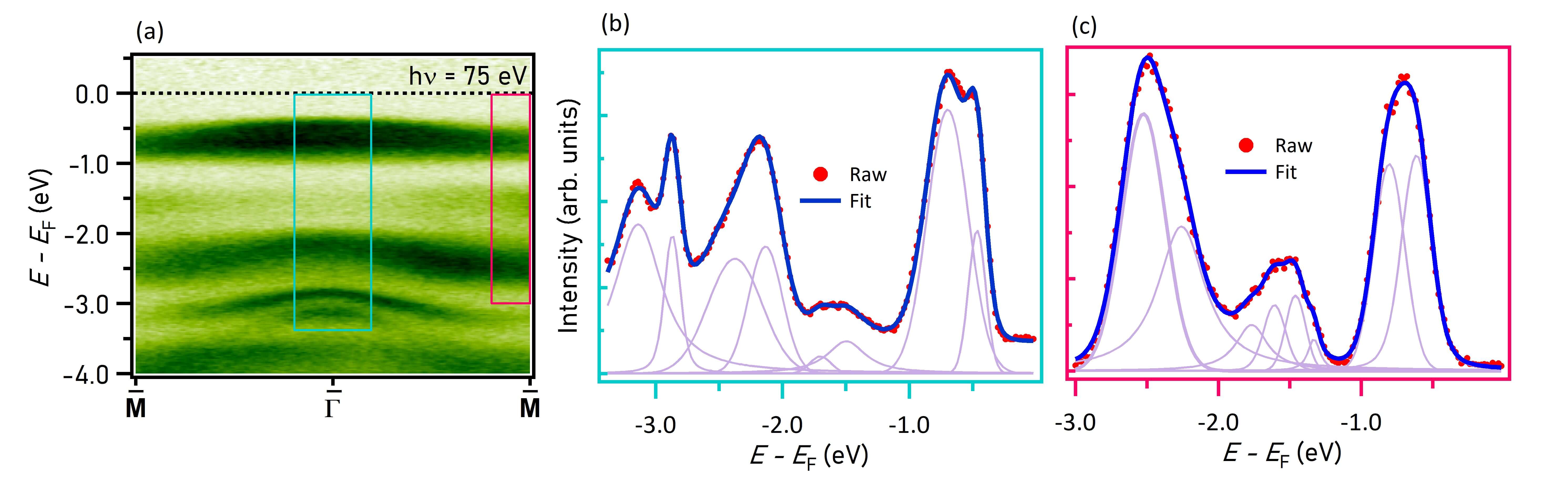}
\caption{$\mathrm{\overline{M}}-\mathrm{\overline{\Gamma}}-\mathrm{\overline{M}}$ band dispersion. (a) Experimental energy-momentum dispersion. (b,c) Integrated EDCs and respective Voigt fits within the cyan and magneta boxes in (a).}
\label{S5}
\end{figure}

\begin{center}
\textbf{4. EXPERIMENTAL BAND DISPERSION ALONG $\mathrm{\overline{M}}-\mathrm{\overline{\Gamma}}-\mathrm{\overline{M}}$}
\end{center}
Experimental electronic band structure along $\mathrm{\overline{M}}-\mathrm{\overline{\Gamma}}-\mathrm{\overline{M}}$ and integrated EDCs taken within the color-coded boxes have been presented in Fig.~\ref{S5}. EDCs have been taken within two regions - one around the center of the Brillouin zone (BZ) and one near the $\mathrm{\overline{M}}$ point. Near $\mathrm{\overline{M}}$, a good fitting is obtained by considering the presence of four close band peaks. These four peaks correspond to two bands in each of the bandsets B and C. Bandsets D and E have very closely spaced bands throughout the BZ, thus identified as a single broad peak in both the regions.

\begin{figure}[h!]
\includegraphics[width=1\textwidth]{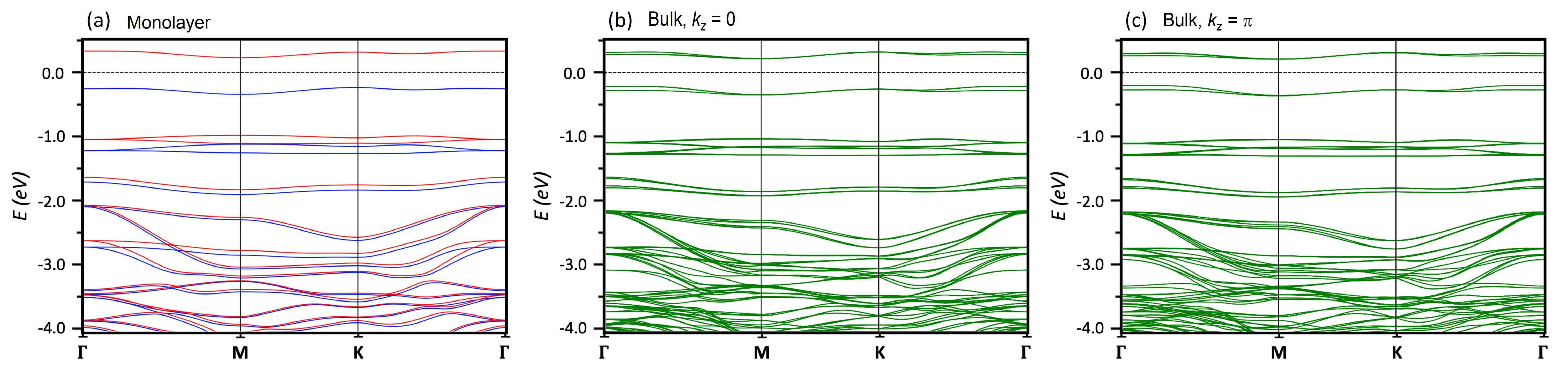}
\caption{Bulk and monolayer band structures. (a) Calculated band structure for monolayer ${\mathrm{Nb}}_{3}{\mathrm{Br}}_{8}$. (b-c) Band structures calculated for bulk ${\mathrm{Nb}}_{3}{\mathrm{Br}}_{8}$ corresponding to $k_z=0$ and $k_z=\pi$ planes, respectively.}
\label{S6}
\end{figure}

\begin{center}
\textbf{5. BAND CALCULATION FOR BULK AND MONOLAYER ${\mathrm{Nb}}_{3}{\mathrm{Br}}_{8}$}
\end{center} 
Fig.~\ref{S6}(a) shows the DFT band structure along various high-symmetry directions for monolayer ${\mathrm{Nb}}_{3}{\mathrm{Br}}_{8}$ and Figs.~\ref{S6}(b,c) show DFT results for the bulk counterpart calculated in different $k_z$ planes as noted on top of each plot. The overall band structure for bulk and monolayer resemble each other, with monolayer having spin-polarized bands because of ferromagnetism. Therefore, the band structure presented in this study for bulk sample is a good representative of how the band structure would look like in the monolayer form. The bulk bands for $k_z=0$ and $k_z=\pi$ are also similar, therefore, making the choice of $k_z$ in calculations not so significant in comparing with the experimental results. 

\begin{figure}[h!]
\includegraphics[width=0.8\textwidth]{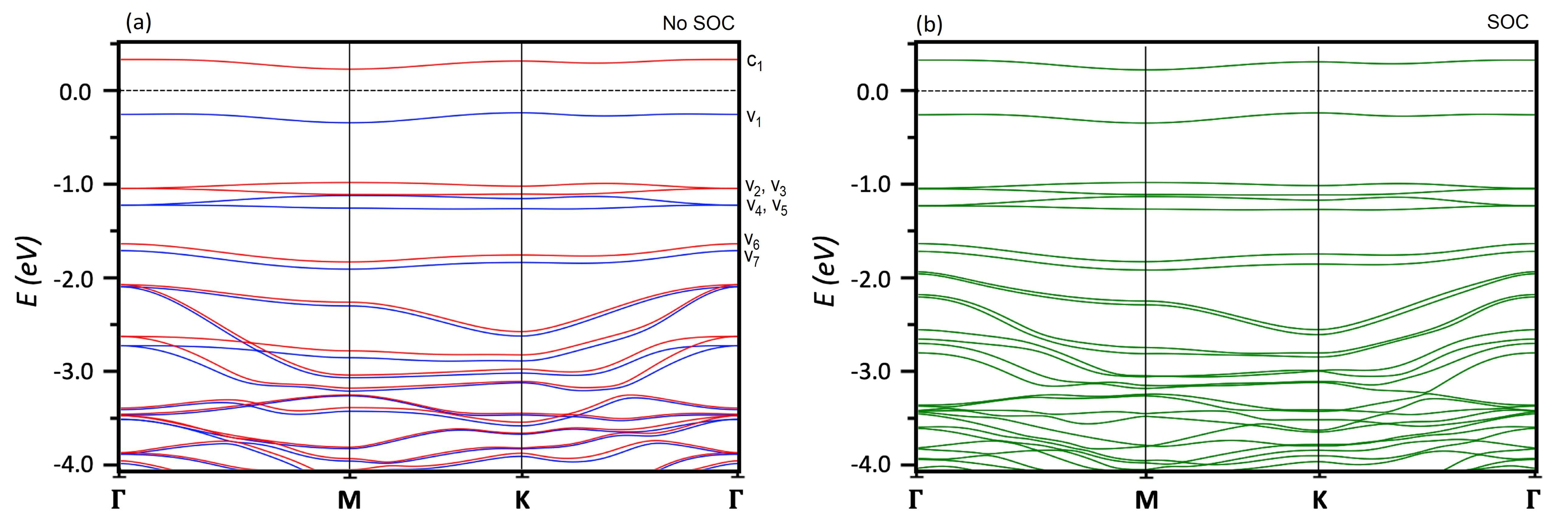}
\caption{Monolayer calculations without (a) and with (b) spin-orbit coupling.}
\label{S7}
\end{figure}

Figure~\ref{S7}(a) shows the monolayer calculations with individual bands labeled c$_1$ and v$_1$ $\rightarrow$ v$_7$ for which we have calculated the mirror operator eigenvalues. The obtained eigenvalues are respectively: 1, 1, 1, -1, 1, -1, 1, 1. The monolayer band structure with the inclusion of spin-orbit coupling is presented in Fig.~\ref{S7}(b).

\begin{figure}[h!]
\includegraphics[width=0.95\textwidth]{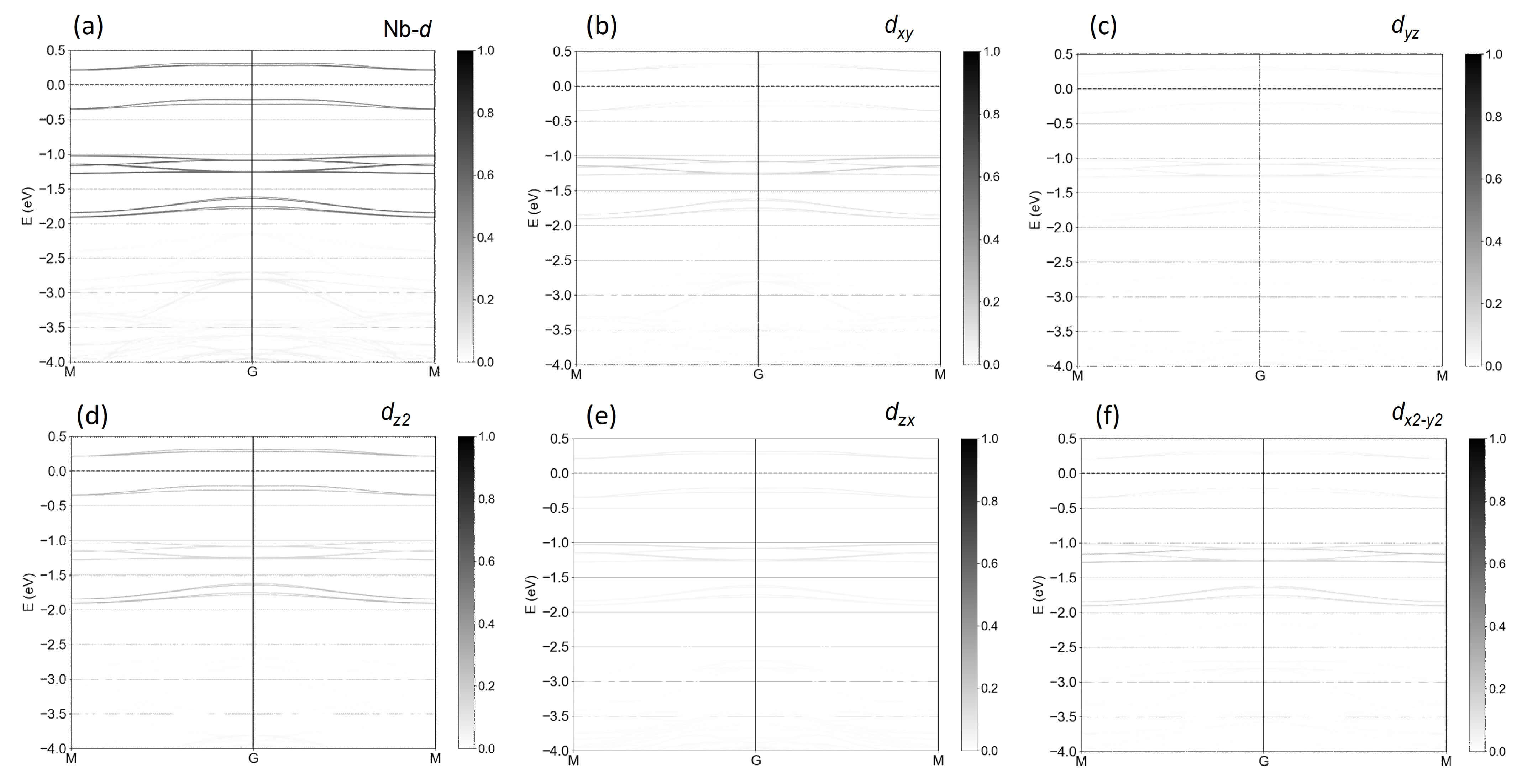}
\caption{Orbital-resolved band structure along $\mathrm{\overline{M}}-\mathrm{\overline{\Gamma}}-\mathrm{\overline{M}}$. (a) Total contribution of $\mathrm{Nb}~d$ in the band structure along the $\mathrm{\overline{M}}-\mathrm{\overline{\Gamma}}-\mathrm{\overline{M}}$ direction. (b-f) Individual contribution of the individual $d$ orbitals of $\mathrm{Nb}$ as marked in each plots.}
\label{S8}
\end{figure}

\begin{center}
\textbf{6. ORBITAL RESOLVED BAND STRUCTURE}
\end{center} 
Figure~\ref{S8} presents the total contribution of $\mathrm{Nb}~d$ orbitals (\ref{S8}(a)) and of individual $d$ orbitals (\ref{S8}(b-f)). The flat and weakly dispersing bands have contribution from $\mathrm{Nb}~d$ orbitals, indicating they are coming from the $\mathrm{Nb}$ atoms, which lie only in the breathing kagome plane. The almost flat and weakly dispersing bands in the bandset A seem to have strong contribution from $d_{z^2}$ and the flat bands in band sets B and C seem to have contribution from all the $d$ orbitals. The relatively dispersive bandsets D and E have strong contribution from  $d_{z^2}$ orbitals along with some contributions from $d_{xy}$ and $d_{x^2-y^2}$.

\end{document}